# Suppression of ferromagnetism in URhGe doped with Ru

S. Sakarya[a], N.H. van Dijk[a], N.T. Huy[b,*] and A. de Visser[b]

[a]*Department of Radiation, Radionuclides & Reactors, Delft University of Technology, Mekelweg 15, 2629 JB Delft, The Netherlands*

[b]*Van der Waals-Zeeman Institute, University of Amsterdam, Valckenierstraat 65, 1018 XE Amsterdam, The Netherlands*

**Abstract**

In the correlated metal URhGe ferromagnetic order ($T_C$ = 9.5 K) and superconductivity ($T_s$ = 0.25 K) coexist at ambient pressure. Here we report on alloying URhGe by Ru, which enables one to tune the Curie temperature to 0 K. URuGe has a paramagnetic ground state and is isostructural to URhGe. We have prepared a series of polycrystalline URh$_{1-x}$Ru$_x$Ge samples over a wide range of $x$ values. Magnetization and electrical resistivity data ($T$ > 2 K) show, after an initial increase, a linear suppression of $T_C$ with increasing $x$. The critical Ru concentration for the suppression of ferromagnetic order is $x_{cr} \approx 0.38$.

The intermetallic compound URhGe attracts much attention because superconductivity ($T_s$ = 0.25 K) coexists with ferromagnetism (Curie temperature $T_C$ = 9.5 K) [1]. The superconducting state is believed to have its origin in the proximity to a ferromagnetic instability: near the quantum critical point enhanced ferromagnetic spin fluctuations mediate Cooper pairing. It is therefore important to study these fluctuations *at* the quantum critical point. Here also the Fermi-liquid theory breaks down. Unlike the case of UGe$_2$ [2], ZrZn$_2$ [3], and UIr [4], hydrostatic pressure (up to 130 kbar) nor uniaxial pressure is effective in tuning URhGe to a quantum critical point, as pressure enhances $T_C$ [5,6]. However, as we will demonstrate $T_C$ can be reduced by chemical substitution.

The choice of elements to substitute on the Rh site with the aim to reduce $T_C$ is limited [7]. UNiGe, UPtGe, and UIrGe are antiferromagnets, UPdGe is a ferromagnet, while UFeGe has a different crystal structure. Since both UCoGe and URuGe are paramagnets and isostructural to URhGe, we decided to investigate the effect of Co and Ru doping. We found that substitution by Co leads to an increase of $T_C$ in the series URh$_{1-x}$Co$_x$Ge ($T_C$ = 20 K for $x$ = 0.60). Here, we will not discuss this series further, but concentrate on the U(Rh,Ru)Ge system. We have carried out magnetization and resistivity measurements on URh$_{1-x}$Ru$_x$Ge samples (0.0 ≤ $x$ ≤ 0.60) and report a complete suppression of ferromagnetism at $x_{cr} \approx 0.38$.

Polycrystalline samples were prepared by arc melting the constituents U, Rh, Ru (all 3N purity) and Ge (5N) under a high-purity argon atmosphere. The as-cast buttons were annealed for ten days at 875 °C. The single-phase nature of the samples was checked by X-ray diffraction and electron microprobe analysis. The magnetization was measured as a function of temperature (1.8 K < $T$ < 20 K) in a magnetic field $B$ = 0.01 T after cooling in zero field. $M(B)$ scans were made in fields up to 5.5 T at several temperatures. Electrical resistivity, $\rho(T)$, measurements were performed using a standard four probe ac technique in zero field down to 2 K.

Our main results are reported in Fig.1, where we show the Curie temperature of URh$_{1-x}$Ru$_x$Ge as a function of Ru concentration $x$, deduced from the magnetic and transport measurements. In the magnetization data $T_C$ was defined as the maximum of -d$M$/d$T$, which agrees well with $T_C$ determined from the Arrott plots extracted from the field scans. In the resistivity data $T_C$ was identified by a pronounced maximum in d$\rho$/d$T$. The values of $T_C$ obtained by both techniques are equal within the experimental error.

---

[*] Corresponding author. Tel.: +31-20-5255627; fax: +31-20-5255788; e-mail: thanh@science.uva.nl (N.T. Huy).

4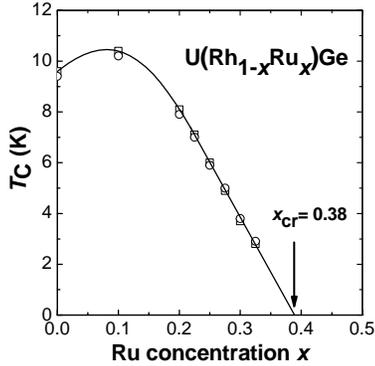

Fig.1 Variation of the Curie temperature $T_C$ with Ru concentration $x$ in URh$_{1-x}$Ru$_x$Ge as determined by magnetization (squares) and electrical resistivity (circles). The solid line serves to guide the eye. The critical Ru concentration for the suppression of ferromagnetic order is $x_{cr} \approx 0.38$.

For our pure URhGe sample $T_C = 9.5$ K, in agreement with literature [1]. Remarkably, when doping small amounts of Ru, the Curie temperature shows an initial increase up to 10.5 K for $x = 0.10$. At higher Ru concentrations $T_C$ is reduced and a linear decrease with a slope of -0.43 K/at.%Ru is observed from at least $x = 0.20$ onwards. For samples with $x = 0.35, 0.375, 0.40$ and $0.60$ we did not detect ferromagnetism in the measured $T$ interval ($T > 1.8$ K). However, the Arrott plots clearly indicate a paramagnetic ground state for $x = 0.40$ and $0.60$. A linear extrapolation of $T_C(x)$ leads to a critical concentration for the suppression of ferromagnetic order $x_{cr} \approx 0.38$.

The suppression of magnetism in correlated metals close to an electronic instability is often discussed in terms of a simple Doniach picture [8], i.e. the competition between the on-site Kondo interaction and inter-site RKKY interaction. In general, pressure tends to suppress magnetism, as the control parameter - the exchange interaction $J$ - increases with increasing hybridization. In the case of URhGe this simple idea does not work, as pressure enhances $T_C$. Consequently, one has to use more sophisticated models, like the one proposed by Sheng and Cooper [9]. By incorporating the change in the $f$-density spectral distribution with space and time under pressure in LMTO band structure calculations, these authors obtain an increase of the magnetic ordering temperature with pressure for compounds like UTe. Such a model could in principle also explain the increase of $T_C$ in URhGe under pressure and the initial increase of $T_C$ in U(Rh,Ru)Ge. However, similar band structure calculations for URhGe have not been performed yet.

It is interesting to compare the volume effect due to Ru alloying and pressure in URhGe. URhGe and URuGe are isostructural, the unit cell volume of URuGe being slightly smaller. X-ray powder diffraction data on some of our samples show that the lattice parameters of the URh$_{1-x}$Ru$_x$Ge series follow Vegard's law. The $a$-axis parameter reduces at a rate of -0.00209 Å/at.%Ru, the $c$-axis parameter increases at a rate of 0.00029 Å/at.%Ru, while the $b$-axis parameter remains constant. With an estimated isothermal compressibility $\kappa = -V^{-1}(dV/dp)$ of 0.8 Mbar$^{-1}$ [5] 10 at.% Ru doping corresponds to a pressure of 3 kbar. With $dT_C/dp = 0.065$ K/kbar [6] this pressure results in a small increase of $T_C$ (9.7 K). Qualitatively this agrees with $T_C \sim 10.5$ K as found for 10 at.% Ru doping. One should however be cautious, because the unit cell reduction upon Ru alloying is strongly anisotropic and dominated by the decrease of the $a$-lattice parameter.

The decrease of $T_C$ beyond $x = 0.10$ is likely attributed to the effect of emptying the $d$-band, since Ru has one electron less than Rh. In a simple model, extracting electrons from the $d$ band results in strengthening the $f$-$d$ hybridization, which in turn leads to a larger exchange parameter $J$ which favours the Kondo interaction. Apparently, this effect dominates the volume effect for $x > 0.10$. We expect disorder to play a secondary role because $T_C$ of URh$_{1-x}$Co$_x$Ge samples with similar disorder gradually increases up to 20 K for $x = 0.60$.

In summary, we have investigated the evolution of ferromagnetism in URhGe alloyed by Ru. $T_C$ initially increases, which is attributed to a volume effect. For x > 0.10 $T_C$ decreases, which suggests that emptying the $d$-band governs the hybridization phenomena. Ferromagnetism is completely suppressed for $x_{cr} \approx 0.38$.

**References**

[1] D. Aoki *et al.*, Nature **413** (2001) 613.
[2] S.S. Saxena *et al.*, Nature **406** (2000) 587.
[3] C. Pfleiderer *et al.* Nature **412** (2001) 58.
[4] T. Akazawa *et al.*, J. Phys.: Cond. Matter **16** (2004) L29.
[5] S. Sakarya *et al.*, Phys. Rev. B **67** (2003) 144407.
[6] F. Hardy *et al.*, Physica B **359-361** (2005) 1111.
[7] V. Sechovsky and L. Havela, in: *Handbook of Magnetic Materials*, Volume 11, Ed. K.H.J. Buschow (Elsevier, Amsterdam, 1998), pp. 1-289.
[8] S. Doniach, Physica B **91** (1977) 213.
[9] Q.G. Sheng and B.R. Cooper, J. Appl. Phys. **75** (1985) 7035.2